\documentclass[graybox]{svmult}

\usepackage{fullpage}
\usepackage{epsfig}
\usepackage{mathptmx}       
\usepackage{helvet}         
\usepackage{courier}        
\usepackage{type1cm}        
\usepackage{makeidx}         
\usepackage{graphicx}        
\usepackage{multicol}        
\usepackage[bottom]{footmisc}
\newcommand{\comment}[1]{}
\newcommand{\x}{{\mathbf x}}

\newcommand{\X}{{\mathbf X}}
\newcommand{\Y}{{\mathbf Y}}
\let\hat\widehat

\makeindex             

\usepackage{amsfonts,amssymb}
\usepackage{natbib}

\begin{document}
\baselineskip 24pt

\title*{Exploiting Non-Linear Structure in Astronomical Data for Improved Statistical Inference}
\titlerunning{Exploiting Non-Linear Structure}
\author{Ann B. Lee and Peter E. Freeman}
\institute{Department of Statistics, Carnegie Mellon University} 
%
%
\maketitle

\abstract{Many estimation problems in astrophysics are highly complex, with high-dimensional, non-standard data objects (e.g., images, spectra, entire distributions, etc.) that are not amenable to formal statistical analysis. To utilize such data and make accurate inferences, it is crucial to transform the data into a simpler, reduced form. Spectral kernel methods 
are non-linear data transformation methods that efficiently reveal the underlying geometry of observable data. Here we focus on one particular technique: diffusion maps or more generally spectral connectivity analysis (SCA). We give examples of applications in astronomy; e.g., photometric redshift estimation, prototype selection for estimation of star formation history, and supernova light curve classification. We outline some computational and statistical challenges that remain, and we discuss some promising future directions for astronomy and data mining.}

\section{Introduction}
\label{sec::Intro}
The recent years have seen a rapid growth in the depth, richness, and scope of astronomical data. This trend is sure to accelerate with the next-generation all-sky surveys (e.g.,  Dark Energy Survey (DES)\footnote{\url{www.darkenergysurvey.org}}, Large Synoptic Survey Telescope (LSST)\footnote{\url{www.lsst.org} \citep{2008arXiv0805.2366I}}, Panoramic Survey Telescope and Rapid Response System (PanSTARRS)\footnote{\url{www.pan-starrs.ifa.hawaii.edu/public}}, Visible and Infrared Survey Telescope for Astronomy (VISTA)\footnote{\url{www.vista.ac.uk}}), hence creating an ever increasing demand on sophisticated statistical methods that can draw fast and accurate inferences from large databases of high-dimensional data.  From a data mining perspective, there are two general challenges one has to face. The first is the obvious {\em computational} challenge of rapidly processing and drawing inferences from massive data sets. The second is the {\em statistical} challenge of drawing accurate inferences from data that are high-dimensional and/or noisy.

\comment{Generally speaking, the performance of statistical estimators (measured, for example, in terms of the mean squared error of an estimated function) degrades quickly as the dimension of the observations (i.e. the number of variables) increases. This problem is often referred to as the statistical ``curse of dimensionality'', and is especially apparent for nonparametric (model-free) methods which make a minimum of assumptions on the distribution of the data.  Any nonparametric estimator of a smooth curve typically has a mean-squared error (MSE) of the form: $ MSE \approx \frac{c}{n^{4/(4+d)}},$ where $n$ is the sample size, $d$ is the dimension of the data, and $c$ is some positive constant. Hence, to ensure a certain precision $MSE=\delta$ we need the number of observations to grow exponentially with the dimension. This means that even if we are able to compute an estimator it may not be accurate.}

Many of the estimation problems in astronomical databases are extremely complex, with observed data that take a form not amenable to analysis via standard methods of statistical inference. 
 To utilize such data, it is  crucial to encode them in a simpler, reduced form. 
 The most obvious strategy is to hand-pick a subset of attributes based on prior domain knowledge. For example, ratios of known emission lines in galaxy spectra may aid in the classification of low-redshift galaxies into starburst, active galactic nuclei, and passive galaxies.  In astrophysical data analysis, a widely used technique for statistical learning is {\em template fitting}, where observed data are compared with sets of simulated or empirical data from systems with known properties; see e.g., \citep{2010MNRAS.403...96B, 2010ApJ...724..425D, 2010ApJ...712..350H, 2010ApJ...708..717S} for some recent template-based work in a variety of astrophysical contexts. Another common data mining approach is 
{\em principal component analysis} (PCA), a globally linear projection method that finds directions of maximum variance; see, e.g., (\citealt{Richards:EtAl:2009a} and references therein; \citealt{Boroson:2010}).


Despite their wide popularity in astrophysical data analysis, the above strategies to statistical learning all have obvious draw-backs: When handpicking a few attributes, one may discard potentially useful information in the data. For template fitting, the final estimates depend strongly on the particular selection of templates as well as the quality of each of the templates. Finally, PCA works best when the data lie in a linear subspace of the high-dimensional observable space, and can perform poorly when this is not the case.

\begin{figure}
\begin{center}
\includegraphics[width=4in]{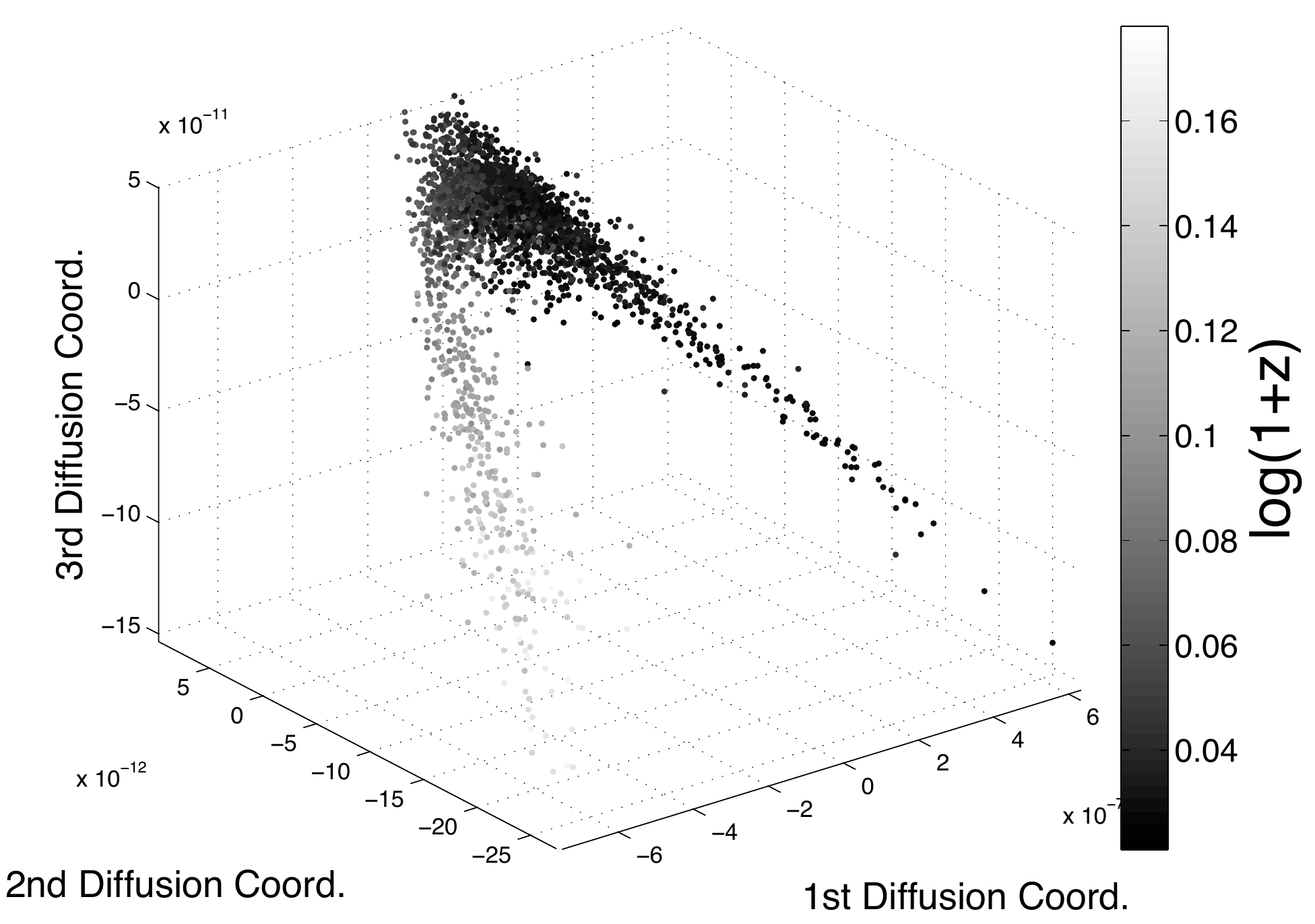}
\end{center}
\caption{\footnotesize 
 Embedding of a sample of 2,793 SDSS galaxy spectra using
        the first 3 diffusion map coordinates.  The gray scale codes for 
        redshift. 
          (Reproduced from~\citealt{Richards:EtAl:2009a})
}
\label{fig::redshift}
\end{figure}

In this paper, we describe a more flexible approach to statistical learning that exploits the intrinsic (possibly non-linear) geometry of observable data with a minimum of assumptions. The idea is that naturally occurring data often have sparse structure due to  constraints in the underlying physical process. In other words, the dimension $d$ of the data space may be large but most of this space is empty. {\em Spectral kernel methods}, such as  spectral clustering~\citep{NgJordanWeiss01, Luxburg:2007}, Laplacian
maps~\citep{BelkinNiyogi03}, Hessian maps~\citep{DonohoGrimes03}, 
and locally linear embeddings~\citep{RoweisSaul00}, analyze the data geometry by using certain differential operators and their corresponding eigenfunctions. These eigenfunctions provide a new coordinate system. For example, consider the emission spectra of astronomical objects. The original data with measurements at thousands of different wavelengths are not in a form amenable to traditional statistical analysis and nonparametric regression. Fig.~\ref{fig::redshift}, however, shows a low-dimensional embedding of a sample of 2,793 SDSS galaxy spectra. The gray scale codes for redshift.  The results indicate that by analyzing 
only a few dominant eigenfunctions of this highly complex data set, one can capture the main variability in redshift, although this quantity was not taken into account in the construction of the embedding. Moreover, the computed eigenfunctions are not only useful coordinates for the data. They form an orthogonal Hilbert basis for smooth functions of the data -- a property that we utilize in~\cite{Richards:EtAl:2009a} for redshift estimation.

More generally, the central goal of spectral kernel methods can be described as follows:
\begin{quote}
{\sf Find a transformation
$Z= \Psi(X)$ such that the structure of the distribution $P_Z$ is simpler
than the structure of the distribution $P_X$
while preserving key geometric properties of $P_X$.}
\end{quote}
``Simpler'' 
can mean lower dimensional but can also be interpreted much more broadly. For example, for redshift prediction using photometric data~\citep{Freeman:EtAl:2009}, we transform the original 16 variables (for magnitude differences between five broad wavelength bandpasses, as
measured using four different magnitude systems) to a 150-dimensional space. For the transformed data, we then fit an additive model of the form
$$ Y = \sum_{i=1}^p \beta_i \psi_i(\x) + \epsilon$$
 where $Y$ denotes observed redshift, $x$ is the original data object (galaxy), $\psi_i(x)$ is the $i$:th coordinate after the transformation, and $\epsilon$ is some random noise.

In this work, we focus on one particular non-linear data transformation called {\em diffusion maps}~\citep{PNAS1,LafonLee2006}, which is an approach to spectral 
connectivity analysis (SCA;~\citealt{Lee:Wasserman:2010}).
SCA analyzes the higher-order connectivity of the data by defining a Markov process on a graph, where each graph node is an observable object, such as a spectrum, galaxy image, or set of light curves for a supernova, etc. The data are then transformed to a metric space where distances reflect the connectivity of the data.  In Sec.~\ref{sec::SCA}, we describe the method. In Sec.~\ref{sec::applications}, we give examples of some applications in astronomy. Finally, in Sec.~\ref{sec::discussion}, we discuss computational and statistical challenges for estimation for large astronomical databases, and outline some promising future directions.



\section{Spectral Connectivity Analysis} \label{sec::SCA}
There  are several data transformation methods that aim to find a 
low-dimensional embedding $Z=\Psi(X)$ of the data while preserving 
key geometric properties of the data distribution $P_X$ in local 
neighborhoods.
Examples of locality-preserving methods are local linear embedding,
Laplacian eigenmaps, 
Hessian eigenmaps, 
local tangent space alignment (LTSA; \citealt{ZhangZha02}),
and diffusion maps.
While the exact details vary, the optimal $r$-dimensional embedding 
(where $r<d$) is provided as the solution to an 
eigenvalue problem, where the
first $r$ eigenvectors $(\psi_1,...,\psi_r)$ provide the new 
data coordinates.

Here we elaborate on diffusion maps  -- a specific approach to spectral connectivity analysis; Euclidean Commute Time maps is a closely related SCA technique discussed in e.g., \cite{FoussEtAl:05}.
Assume we observe data $\mathcal{X}_{\rm obs}=\{\x_1, \ldots, \x_n\}$, where $\x \in \mathbb{R}^d$.  
 The basic idea is to create a distance 
$D(\x_i,\x_j)$  that measures ``connectivity'' or how 
easily information ``flows'' from point $\x_i$ to $\x_j$ in a Markov chain 
on the observed data.
(The data ``points" $\x_i$ and $\x_j$ represent {\em entire} observable objects; for example,  the full emission spectra of two astronomical objects, images of two galaxies, or  light curves of two supernovae; $D$ is a measure of distance between the objects.)
High flow occurs in high-density regions, and points that are connected 
by many high-flow paths are close with respect to the diffusion metric.
The transition matrix $A$ of the Markov chain is based on a 
user-defined pairwise distance $\Delta(\cdot,\cdot)$ that is a good measure of dissimilarity in local neighborhoods; a common choice is the 
Euclidean distance in $\mathbb{R}^d$ but other dissimilarity measures that incorporate prior knowledge and measurement errors can also be used. We define the transition probability from $\x_i$ to $\x_j$ in one step by $A(\x_i,\x_j)=\frac{\exp(-\Delta(\x_i,\x_j)/\epsilon)}{\sum_k \exp(-\Delta(\x_i,\x_k)/\epsilon)}$, where $\epsilon>0$ is a tuning parameter that determines the local neighborhood size. Let $A_t(\x_i,\x_j)$ denote the $t$-step transition probability; the parameter $t$ determines the amount of smoothing along high-density regions and the ``scale'' of the analysis. The diffusion distance between points $\x_i$ and $\x_j$ is defined as
\begin{equation}
D(\x_i,\x_j) =  \sum_{\mathbf{z} \in \mathcal{X}_{\rm obs}} \frac{(A_t(\x_i,\mathbf{z})-A_t(\x_j, \mathbf{z}))^2}{\phi_0(\mathbf{z})} ,
\end{equation}
where the sum is over all points $\mathbf{z}$ in the data set $\mathcal{X}_{\rm obs}$, 
and $\phi_0(\mathbf{z})$ is the stationary distribution of the Markov chain as $t \rightarrow \infty$. 
In practice, we never explicitly implement the
Markov chain but instead solve an eigenproblem for an n-by-n matrix.
Let $\lambda_1 \geq \lambda_2 \geq \ldots$ and $\{\psi_i\}$ be the 
eigenvalues and corresponding right eigenvectors of the 1-step transition matrix $A$.
The diffusion map $\Psi_t\!: \mathcal{X}_{\rm obs} \rightarrow \mathbb{R}^r$  (where $r<n$) 
is given by
\begin{equation}
\Psi_t(\x) = (\lambda_1^t \psi_1(\x), \ldots, \lambda_r^t \psi_r(\x)) .
\label{eq:diffmap}
\end{equation}
As shown in~\cite{Coifman:Lafon:06} and Lafon \& Lee (2006), it holds that 
$D_t(\x_i,\x_j) \approx \|\Psi_t(\x_i)-\Psi_t(\x_j) \|$,
i.e., the Euclidean distance in the new coordinate system approximates 
the diffusion distance in the original coordinate system.
Because all connections between data points are
simultaneously considered, diffusion maps are
robust to noise and outliers and they return embeddings where
metrics are explicitly defined.

Incorporating data geometry via $\Psi$ and SCA can lead to radically
improved inference algorithms. For details on the statistical properties of SCA refer to~\cite{Lee:Wasserman:2010}.  In Sec.~\ref{sec::applications}, we give examples of some specific applications in astronomy. 


\subsubsection*{Out-of-Sample Extensions of Empirical Data Sets}
 Let $\mathcal{X}$ denote the space of all data. One can show that the random walk and the eigenvectors $\{{\psi_j}\}$ derived from the finite set $\mathcal{X}_{\rm obs}$
 have meaningful limits as the sample size $n \rightarrow \infty$. 
 Hence, we can think of the eigenvectors  of the discrete random walk  as estimates  of eigenfunctions $\{{\psi_j(\x)}\}_{j \in \mathbb{N}}$, defined on $\mathcal{X}$,  at the {\em observed} values $\x_1,\ldots, \x_n$. We estimate the function $\psi_j(\x)$ 
at values of $\x$ not corresponding to one of the $\x_i$'s in the data set by 
the kernel-smoothed estimate
\begin{equation}\label{eq::nystrom}
\hat\psi_{j}(\x)=
\frac{1}{\lambda_{j}} \sum_{i=1}^{n} A(\x, \x_i) \psi_j(\x_i),
\end{equation}
where $A(\x,\x_j)=\frac{\exp(-\Delta(\x,\x_j)/\epsilon)}{\sum_k \exp(-\Delta(\x,\x_k)/\epsilon)}.$
This expression is known in the 
applied mathematics literature as the
Nystr\"{o}m approximation. These out-of-sample extensions allow us to make predictions for new data points that are not in the sample using diffusion maps and, for example, adaptive regression (as in Sec.~\ref{sec:adaptive_regression}).


\section{Applications in Astronomy}\label{sec::applications}
In this section, we give some examples of applications of SCA to astrophysical problems. Among other things, diffusion maps can be used to estimate parameters in a regression framework, build classification models, and select prototypes for parameter estimation in complex models. The details are described in separate papers.

\subsection{Adaptive Regression and Redshift Estimation}\label{sec:adaptive_regression}
In~\cite{Richards:EtAl:2009a}, we show how one can take advantage of the underlying data structure in non-parametric regression such as redshift prediction. The main idea is to describe the intrinsic data geometry in terms of fundamental eigenmodes. These eigenmodes can be viewed both as (i) {\em coordinates} of the data, as in Fig.~\ref{fig::redshift}, and as (ii) {\em orthogonal basis functions} for curve estimation. The latter insight can be used to develop a general regression framework for sparse, complex data.


 Let  $\mathcal{X} \subset \mathbb{R}^d$ denote the space of all observed data.  In regression, the goal is to predict a real-valued function $f(\mathbf{x})$ for data $\mathbf{x} \in \mathcal{X} $, when given a sample of known pairs $(\mathbf{x},Y)$ where the response $Y = f(\mathbf{x}) + \epsilon $. If $f \in L^2(\mathcal{X})$ and $\{\psi_1,\psi_2,\ldots,\}$ is an orthonormal basis, then we can write
$$ f(\mathbf{x}) = \sum_{j=1}^{\infty} \beta_j \psi_j(\x),$$
where the expansion coefficients $\beta_j = \int f(\x) \psi_j(\x) d\x$. The standard approach in non-parametric curve estimation~\citep{efromovich1999nonparametric} is to choose a fixed known basis (e.g., Fourier or wavelet bases) for, for example, $L^2([0,1])$, and then extend the basis to two or three dimensions by a tensor product. Such an approach quickly becomes intractable in higher dimensions. In astrophysical problems, the response $Y$ may be the redshift, age or metallicity of galaxies, and $\x$ is often a high-dimensional, non-standard data object, such as the emission spectrum measured at $p>1000$ wavelength bins, or photometry data in a color space with $p>10$ dimensions.

In~\cite{Richards:EtAl:2011a}, we suggest a new, adaptive approach to non-parametric curve estimation, which utilize the data-driven (orthogonal) eigenfunctions $\{\psi_1,\psi_2,\ldots,\}$ computed by PCA or  spectral kernel methods. The regression function estimate $\hat{r}(\x)$ is then given by 
$$ \hat{f}(\x)  = \sum_{j=1}^{p} \hat{\beta}_j \psi_j(\x),$$
where the coefficients $\hat{\beta}_j$ are estimated from the data, and $p$ is a smoothing parameter determined by cross-validation and a mean-squared error prediction risk.   The method is computationally efficient, making it appropriate for large databases such as the SDSS. One can use the predictions to speed up more computationally expensive estimation techniques by narrowing down the relevant parameter space; e.g., the redshift range or the set of templates in cross-correlation techniques. Adaptive regression also provides a useful tool for quickly identifying  outliers; e.g., misclassified spectra, spectra with anomalous features, etc. In Richards et al., we consider a sample of 3835 galaxy spectra from the SDSS database.  For this data, the estimates based on eigenmodes from SCA (diffusion maps) lead to markedly better predictions than estimates from PCA, indicating the importance of non-linear geometries.

The development of fast and accurate methods of photometric redshift estimation is a crucial step towards being able to fully utilize the data of next-generation surveys for  precision cosmology. In~\cite{Freeman:EtAl:2009}, we apply adaptive regression and SCA to the problem of {\em photometric redshift estimation} for three different data sets: 350,738 SDSS main sample galaxies, 29,816 SDSS luminous red galaxies, and 5,223 galaxies from DEEP2 with CFHTLS $ugriz$ photometry.  For computational speed, we first derive diffusion coordinates for training sets limited to about 10$^4$ galaxies, and then extend these coordinates to the full data sets by the Nystr\"{o}m method. 
The final redshift predictions achieve an accuracy on par with that of existing ML-based techniques, e.g., artificial neural networks~\citep{2004PASP..116..345C} and $k$-nearest neighbors~\citep{2008ApJ...683...12B}.

\subsection{Prototype Selection for Estimation of Star Formation History}
Parameter estimation in astronomy and cosmology often requires the use of complex physical models. In a typical application the mapping from the parameter space to the observed data space is built on sophisticated physical theory or simulation models or both. 
\comment{Here is one such scenario: Imagine we have a computer model that can generate clean, idealized versions 
 of theoretical data with known parameters.
 The actual observations,
  however, are very complex and noisy.  In some cases, they can be appropriately modeled as linear combinations of the theoretical simpler components, subject to non-linear transformations and observational noise. The physical parameters of interest are estimated by fitting this model to observed data. An important question is: How should one choose the set of prototypes? A smaller set will make the computations faster, but a principled choice of examples can, as we shall see, also lead to more accurate parameter estimates.}
In ~\cite{Richards:EtAl:2009b,Richards:EtAl:2011a}, we study one such scenario: the problem of estimating star formation history (SFH) in galaxies given SDSS high-resolution spectra. A common technique in the astronomy literature, called {\em empirical population synthesis} (see e.g., \citealt{2001MNRAS.325...60C} and references within), is to model each galaxy as a mixture of stars from different simple stellar populations (SSPs), where an SSP is defined as a group of stars with the same age and metallicity.
 The principle behind this method is that each galaxy consists of multiple subpopulations of stars of different ages and compositions so that the integrated observed light from each galaxy is a mixture of the light contributed by each SSP.  By estimating the mixture coefficient of each SSP, one can then reconstruct the star formation rate and composition as a function of time throughout the life of that galaxy. 

\comment{Describing the data from each galaxy as a combination of SSPs allows us to reconstruct the star formation and metallicity history of each galaxy.   This is because, for each galaxy, the component weight on an SSP captures the proportion of that galaxy's stars that was created at the specific epoch corresponding to the age of that SSP.  Therefore, the full vector of SSP component weights for each galaxy describes the star formation throughout the galaxy's lifetime.}

In our work, we use theoretical SSP models by \cite{BruzualCharlot:2003}. For the galaxy spectra, we adopt the empirical population synthesis model in \cite{CF2004}: 
\begin{equation}
  \label{cfmodel}
  \Y_{\lambda} (\mathbf{\gamma}, M_{\lambda_0}, A_V, v_*, \sigma_*)=
  M_{\lambda_0}\left(\sum_{j=1}^{N}\gamma_j \X_{j,\lambda} r_{\lambda}(A_V)\right) \otimes G(v_*,\sigma_*)
\end{equation}
where $\Y_{\lambda}$ is the light flux at wavelength $\lambda$; $\X_{j}$ is the normalized $j$th SSP spectrum; and $\gamma_j \in [0,1]$ is the proportion of luminosity contributed by the $j$th SSP. (The remaining model parameters describe the flux normalization and observational noise, such as the amount of reddening due to foreground dust, spectral distortions due the movement of stars within the observed galaxy, etc.) We fit the signal model in Eq.~\ref{cfmodel} to observed noisy galaxy data with maximum likelihood estimation and MCMC.  We then derive various physical parameters of interest from the SSP parameters (which are known) and the component weights in the signal model (which are estimated). For example: the average log age of the stars in a galaxy, $\langle \log t \rangle = \sum_{j=1}^N \gamma_j \log t(\X_j)$, where $t(\X_j)$ is the age of the $j$:th SSP; similarly, the average log metallicity $\langle \log Z \rangle = \sum_{j=1}^N \gamma_j \log Z(\X_j)$, where $Z(\X_j)$ is the metallicity of the $j$:th SSP.

\begin{figure}
  \centering
  \includegraphics[width=4in]{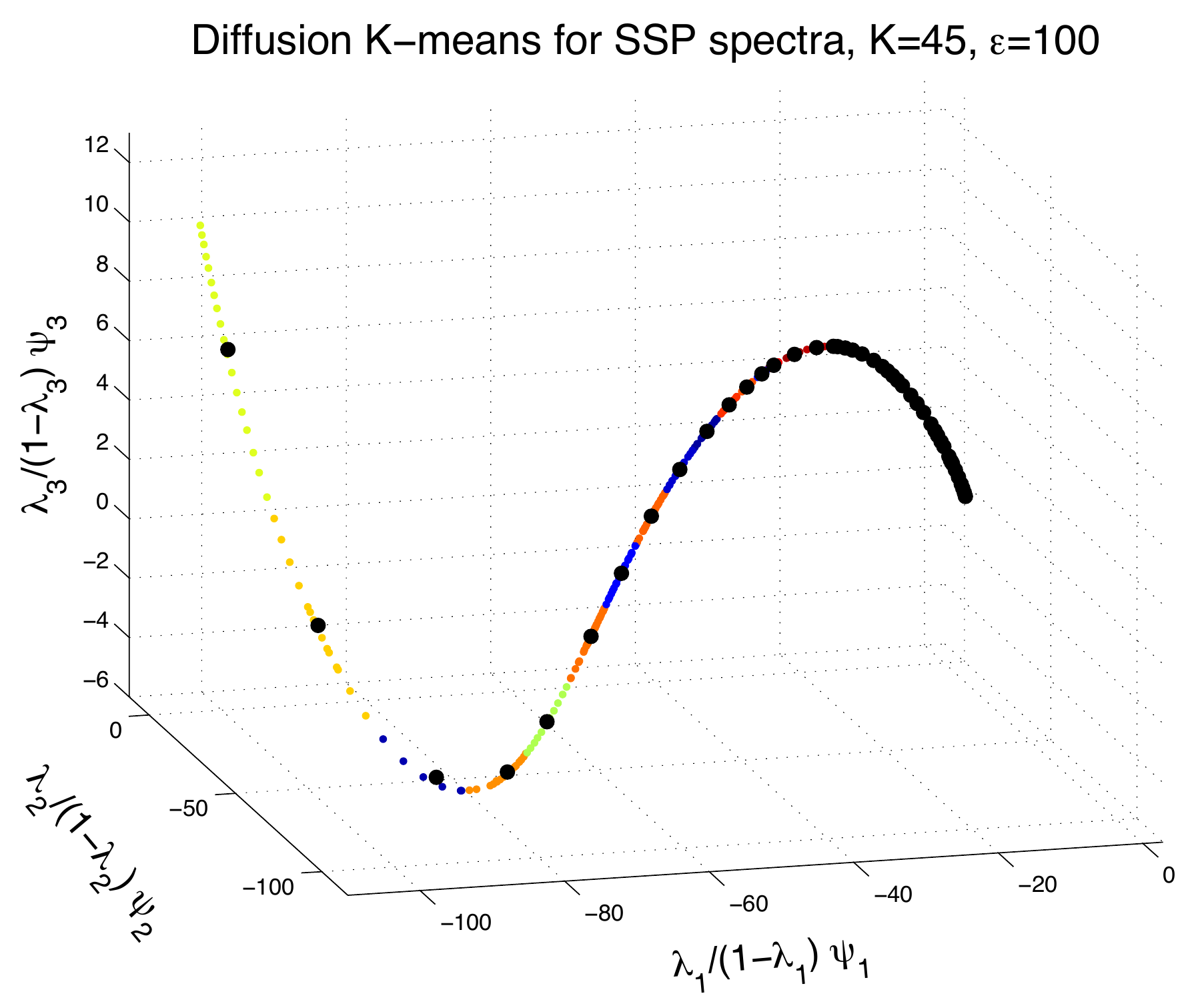}
  \caption{\footnotesize{Prototyping of SSP spectra by Diffusion $K$-means.}  Representation of
    1278 SSP spectra in 3-dimensional diffusion space.  Large black
    dots denote the $K=45$ centroids.  Individual SSPs are colored by
    cluster membership.  The theoretical SSPs reside on a simple, low-dimensional
    manifold which is captured by the $K$ prototypes. (Reproduced from~\citealt{Richards:EtAl:2009b})}
  \label{sspdmap}
\end{figure}

\begin{figure}
  \centering
\includegraphics[width=4in]{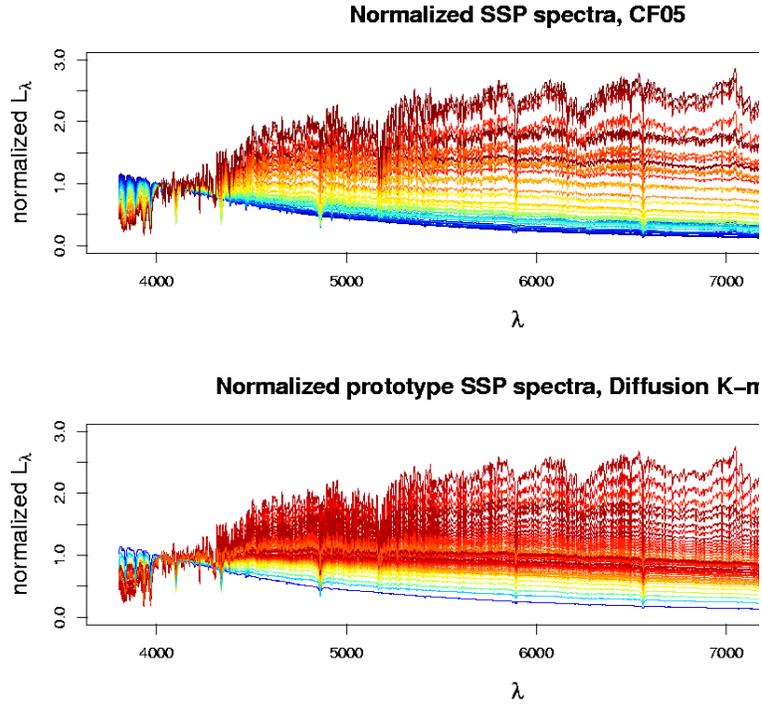}
  \caption{\footnotesize{Basis spectra for CF05 and Diffusion $K$-means,
      colored by $\log t$.}  All
  spectra are normalized to 1 at $\lambda_0=4020$ \AA.  The diffusion
  $K$-means basis covers the spectral range in relatively uniform increments,  while the CF05 basis oversamples spectra from young stellar populations and undersamples the spectral range of older populations. (Reproduced from~\citealt{Richards:EtAl:2009b})}
  \label{protot}
\end{figure}

An important question is: How should one choose the set of SSPs? Though the parameters that define each SSP are continuous, optimizing the signal model over a large set of SSPs on a fine parameter grid is computationally infeasible and inefficient. As we shall see, it also leads to poor statistical estimates. In \cite{Richards:EtAl:2011a}, we introduce a principled approach of choosing a small basis of SSP \emph{prototypes} for optimal SFH parameter estimation.  The basic idea is to explore the underlying geometry of the SSP observable data, and quantize the vector space and effective support of these model components.  In addition to greater computational efficiency, we achieve better estimates of the SFH target parameters. In simulations, our proposed quantization method obtains a substantial improvement in estimating the target parameters over the common method of employing a parameter grid.
The main reason for the improvement is that under the presence of noise, components with similar 
 functional forms will be indistinguishable. Hence, it is more advantageous to choose prototypes that are approximately evenly spaced in the space of model data rather than evenly spaced in the parameter space.  By replacing the theoretical models in each neighborhood by their local average in diffusion space (``Diffusion $K$-means''; Figure~\ref{sspdmap}), the model quantization approach is optimal for treating degeneracies because it allows a slight increase in bias to achieve a large decrease in variance of the target parameter estimates.  See Figure~\ref{protot} for a plot of two SSP spectral bases with $K$ prototypes chosen by a regular parameter grid and by our proposed quantization method, respectively.

\subsection{Supernova Classification}

In many astronomical problems, classification is of paramount importance.
For instance, one may be interested in determining which of a collection of
light curves is associated with RR Lyrae stars, or Cepheids, etc.
Depending on the problem, classification may be done in an unsupervised manner, to
uncover hidden structure in the data, or, if at least some of the data labels
are known, a classifier can be trained and then used to predict the classes
of unlabeled data.

The next generation of survey telescopes will observe hundreds of thousands of noisy 
and irregular photometric SN light curves, from which astronomers will want
to construct highly pure and efficient Type Ia SN samples for use in testing cosmological
theories.  In \cite{Richards:EtAl:2011b}, we apply a semi-supervised approach to supernova
classification.  In the unsupervised step, we fit regression splines to each of a set of light curves, then
via diffusion map place them in a lower-dimensional embedding space that capture the geometry of the underlying data distribution.  In that space,
we then take the supervised step of training a random forest classifier with only the labeled data, 
with the results used to classify the unlabeled data.  Applied to the data of the 
Supernova Photometric Classification Challenge (Kessler et al. 2010), we achieve 96\%
purity and 86\% efficiency when labeling the training set; for the test set, the figures
are 56\% and 48\% respectively.  As the sample sizes (of unlabeled and/or labeled data) increase, our 
semi-supervised approach will yield progressively more accurate classifications, in contrast to template-based approaches which do not benefit from larger data sets.  We also explore how different spectroscopic followup strategies
affect these figures, finding that deeper surveys yielding fewer labeled SNe can produce better
results than shallower surveys.  Determination of an optimal labeling strategy
is an important component of {\em active learning}, a
topic we will return to in the discussion below. 

\section{Discussion and Future Directions}\label{sec::discussion}

In this review, we have described SCA --- a statistical technique for transforming  complex, data objects into a coordinate system that reveals the structure of the underlying data distribution. Such a transformation may improve the performance of classification, regression, clustering and parameter estimation. We have seen applications of SCA in redshift prediction, estimation of star formation history and photometric supernova classification. Currently, we are working with Chad Schafer to develop SCA as a tool for combining theoretical modeling and observational evidence into optimal constraints on the parameters of physical models. The idea is to map observed data (e.g., light curves of Type Ia supernova) as well as distributions for the observable data, constrained by physical theory (e.g., cosmological models) into a simpler encoding space. The shared representation of data and distributions is then exploited to achieve optimal constraints on physical theories, in the form of set estimators on the distribution space; see Schafer's SCMA 2011 talk and paper for details.

Another promising direction of SCA is {\em semi-supervised learning} (SSL), in particular, in combination with {\em active learning}. Suppose that we have a regression or classification problem. The typical scenario in SSL is that we have access to a large database of unlabeled examples (e.g., photometric data with unknown redshift), but relatively few labeled examples (e.g., data with spectroscopically confirmed redshift). Classical regression and classification techniques only take advantage of labeled data, but the central idea behind SSL is that one can make use of the unlabeled data to improve predictions; see e.g., \cite{Belkin:Niyogi:04, Lafferty:Wasserman:07, Singh:08} for theoretical results on SSL.  In our supernova classification application, we showed that learning a low-dimensional coordinate system using unlabeled data improves subsequent classification by trees. We also found evidence that the exact choice of training examples has a large effect on the results. In future work, we plan to explore whether we can achieve greater accuracy in classification and regression problems with fewer training labels if a so-called active learner is allowed to repeatedly pose {\em queries}, in the form of unlabeled data instances to be labeled by an oracle. In the machine learning literature, there are many variants of active learning; see, e.g., \cite{Settles:2010} for a literature survey. All these models involve a search through the hypothesis space. Such searches and subsequent queries could potentially be better adapted to the underlying data distribution via an unsupervised technique such as SCA that exploit clusters and groupings in data.

Finally, there are the computational challenges of efficiently constructing weighted graphs and performing eigencomputations for very large databases.
 We are currently exploring several solutions --
 most notably, fast approximate nearest neighborhood searches via trees, eigencomputations via streaming PCA~\citep{2009MNRAS.394.1496B}, very large-scale algebraic computations via matrix randomization~\citep{Halko:EtAl:2011}, and subsampling combined with Nystr\"om extensions to reduce the size of the distance matrix that is effectively eigendecomposed.

\section{Acknowledgments} Part of this work is joint with Joseph W. Richards, Chad M. Schafer, Jeffrey A. Newman, and Darren W. Homrighausen. We would also like to acknowledge ONR grant \#00424143, NSF grant \#0707059, and NASA
AISR grant NNX09AK59G.


\bibliographystyle{chicago}

\end{document}